\documentclass[twocolumn,twoside]{article}
\usepackage{amssymb}
\usepackage{float}
	\usepackage{amsmath}
	\usepackage{abstract}
	\usepackage{graphicx}
	\usepackage{authblk}
	\usepackage{dcolumn}
	\usepackage{bm}
	\usepackage{inputenc}
	\usepackage{color}
	\usepackage{authblk}
	\usepackage{breqn}
	\usepackage[square,sort,comma,numbers]{natbib}
	\usepackage{url}
	\usepackage{geometry}
\title{Hollow Bessel beams for guiding atoms between vacuum chambers: A proposal and efficiency study}

\author[1]{D. Rivero}
\author[2]{V.S. de Angelis}
\author[1]{C.Beli}
\author[1]{M. Moreno}
\author[2]{L. A. Ambrosio}
\author[1,*]{Ph. W. Courteille}
\affil[1]{São Carlos Institute of Physics, University of São Paulo, 13560-970 São Carlos, SP, Brazil}
\affil[2]{São Carlos School of Engineering, University of São Paulo, 13560-970 São Carlos, SP, Brazil}
\affil[*]{philippe.courteille@ifsc.usp.br}
\date{\vspace{-5ex}}

\begin{document}
\twocolumn[
\begin{@twocolumnfalse}
\maketitle 
\begin{abstract}
We explore a scheme for guiding cold atoms through a hollow Bessel beam generated by a single axicon and a lens from a 2D magneto-optical trap toward a science chamber. We compare the Bessel beam profiles measured along the optical axis to a numerical propagation of the beam's wavefront, and we show how it is affected by diffraction during the passage through a long narrow funnel serving as a differential pumping tube between the chambers. We derive an approximate analytic expression for the intensity distribution of the Bessel beam and the dipolar optical force acting on the atoms. By a Monte-Carlo simulation based on a stochastic Runge-Kutta algorithm of the motion of atoms initially prepared at a given temperature we show that a considerable enhancement of the transfer efficiency can be expected in the presence of a sufficiently intense Bessel beam.
\end{abstract}
\url{https://doi.org/10.1364/JOSAB.395200}
\vspace{.7cm}
\end{@twocolumnfalse}
]

\section{Introduction}

Experiments with ultracold atoms are conditioned to the availability of an extreme ultra-high vacuum (XUHV) environment. This entailed the development of a variety of techniques allowing for a physical separation of a science chamber, maintained at pressures as low as $10^{-11}~$mbar or below, and a preparation chamber. In the preparation chamber the atomic gas is provided by heating a solid chunk of a specific element (in many cases alkaline or alkaline-earth metals) to temperatures sufficiently high to reach noticeable partial pressures of typically above $10^{-9}~$mbar. A long narrow pipe connects both chambers and maintains a differential vacuum. The most common techniques for transferring the atoms between the chambers are the Zeeman slower \cite{Phillips82,Prodan85,Bagnato87}, the double magneto-optical trap (MOT) \cite{Myatt96,Prevedelli99}, the 2D-MOT \cite{Dieckmann98}, and different types of conveyor belt techniques \cite{Barrett01,Haensel01,Fortagh98}.

The transfer of atoms is always subject to atomic losses due to atoms escaping from the beam. Some experiments resort to funneling the atoms through quadrupolar or hexapolar magnetic waveguides \cite{Myatt96,Prevedelli99}, which however only works for atoms that are paramagnetic in their ground state. In this paper, we study the idea of guiding atoms through a hollow Bessel beam (BB) over a distance of 23~cm, from a 2D-MOT created in a preparation chamber, through a 20~mm long and 2~mm wide differential vacuum tube, to a distant 3D-MOT operated in a science chamber. If the frequency of the BB is tuned to the blue of an atomic resonance, the repulsive force exerted on the atoms by the dipolar optical potential will keep the atoms away from the walls of the tube.

Hollow beams in free space, such as Laguerre-Gaussian beams or BBs have been proposed for guiding or trapping atoms \cite{Manek98,SongY99,Kulin01,XinyeXu01,Arlt02,Friedman02,Rhodes02,Rhodes06,ZhaoyingWang05,Fatemi06,Mestre10,Carrat14}. In some cases atoms were guided through the hollow core of a light-carrying optical fiber \cite{Bajcsy11,Poulin11,Pechkis12}. For instance, Song et al.~\cite{SongY99}.~prepared a cloud of $100~\mu$K cold cesium atoms from a MOT directly inside the hollow beam tuned 0.25..1.5~GHz to the blue of the Cs cooling transition. They observed acceleration and heating of the atoms due to residual Rayleigh scattering. Xinye Xu et al.~\cite{XinyeXu01}demonstrated a 20-fold increase of the transport efficiency of atomic rubidium over a 10~cm long distance through the hollow beam produced by a hollow-core optical fiber in good agreement with Fokker-Planck simulations. And Carrat et al.~\cite{Carrat14} meticulously characterized the channeling of cold rubidium atoms prepared in a
2D-MOT through a Laguerre-Gauss beam generated by a spatial light modulator. Most experiments rely on Laguerre-Gaussian laser beams, which are easy to produce and stay hollow over very long distances. In this paper we propose and analyze a scheme for guiding a precooled cloud of strontium-88 atoms from a preparation chamber through a narrow funnel into a science chamber via a hollow BB.

Bessel beams are characterized by extremely localized tube-like intensity distribution walls, giving rise to very deep dipolar potentials for atoms. On the other hand, they maintain their shape only over limited distances.

The paper is organized in two main parts: First, we propose a geometry of optical components for creating a hollow BB using a single axicon and a lens. Numerically simulating the evolution of the intensity distribution along the optical axis and mapping it out experimentally, we find that the hollow BB stays diffraction-free over a distance of more than 20~cm. Second we present an analytical approximation of the intensity distribution, which will turn out to be useful for the Monte-Carlo simulations of atomic trajectories presented in the second part of the paper.

\begin{figure}[h!]
    \centerline{\includegraphics[width=8.7cm]{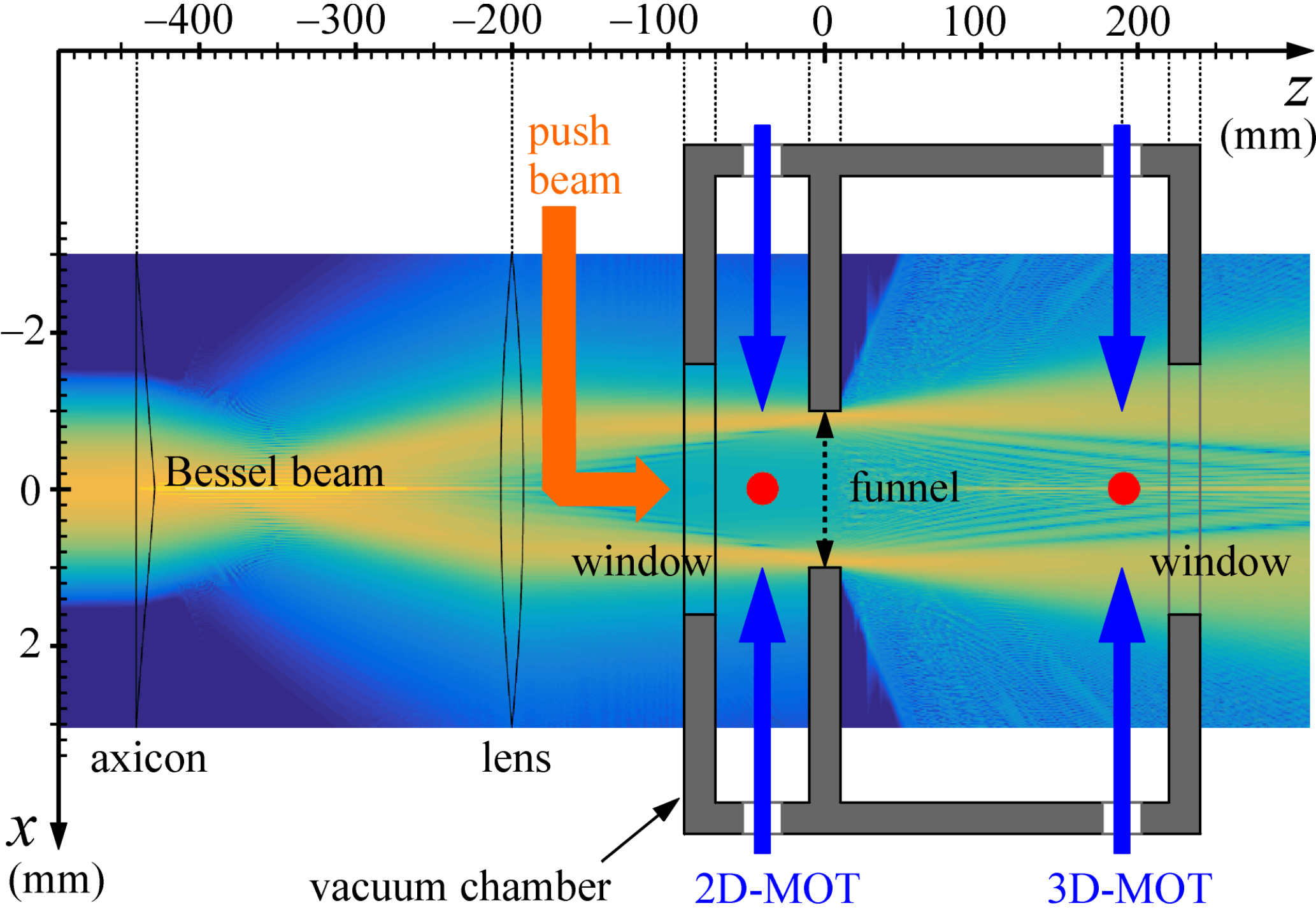}}	
    \caption{(color online, not to scale) Scheme of the proposed atomic guiding: A push beam and a Bessel beam are injected collinearly to the optical axis through a vacuum viewport onto a 2D-MOT, where they create a guided atomic beam. The laser beams and the atomic beam are funneled through a 20~mm long and 2~mm wide differential pumping tube before they cross the capture region of the 3D-MOT.}
    \label{fig:Scheme}
\end{figure}

\section{Bessel beams}

Ideally, a BB is a non-diffracting monochromatic solution to the scalar wave equation in cylindrical coordinates carrying an infinite amount of energy \cite{Durnin87,McGloin05}. In its simplest form, the electric field of an arbitrary $\nu$-th order BB with wavelength $\lambda$ can be written as,
\begin{equation}
	\label{eq:Bessel01}
	E(\rho,\phi,z) = A_0\exp(ik_zz)J_\nu(k_\rho\rho)\exp(i\nu\phi)~,
\end{equation}
 where $A_0$ is the electric field strength and $J_\nu$ is the $\nu$-th order Bessel function of the first kind. In Eq. (\ref{eq:Bessel01}), $k_z$ and $k_\rho$ are the longitudinal and transverse wave numbers satisfying the dispersion relation $k^2=k_z^2+k_\rho^2=(2\pi/\lambda)^2$, such that $k_z=k\cos\theta$ and $k_\rho=k\sin\theta$, being $\theta$ the axicon angle associated to the tilted plane of waves propagating along the surface of a cone of half-angle $\theta$ in the angular spectrum decomposition. Cylindrical coordinates $(\rho,\phi,z)$ have been adopted, and a time harmonic factor $\exp(-i\omega t)$ has been omitted for brevity. For our purposes, the Rayleigh range is an essential parameter to be considered since the non-diffracting beam must propagate through a 20~cm long differential vacuum tube to minimize losses of atoms during their guidance to the science chamber. The maximum propagation distance up to which a BB can overcome diffraction is given by $Z_{max}=2\pi R\bar r/\lambda$, where $R$ is the aperture radius and $\bar r$ is the beam radius \cite{Durnin87}. It should be noticed that, in general, $Z_{max}$ is much greater than the Rayleigh range of a Gaussian beam with an equivalent beam waist radius $w_g=\bar r$.

\subsection{Numerical propagation of the phase front}

In practice, various techniques have been developed for generating BBs, the simplest one being based on axicons \cite{McLeod54,Depret02,Tsai06,Brzobohaty08}. In this section we numerically simulate the evolution of the phase front of a Gaussian laser beam passing through an axicon and then a focusing lens. 

To propagate an optical phase front located at the position $z_0$ of the optical axis, we conveniently start from a Gaussian laser beam of waist radius $w_g$ and electric field strength $B_0$,
\begin{equation}
	\label{eq:Bessel11}
	E_{Gauss}(x,y,z_0) = B_0\exp(-\rho^2/w_g^2)~,
\end{equation}
where $\rho=\sqrt{x^2+y^2}$ is the distance from the optical axis. The electric field is normalized to the total power $P$ of the light beam, that is, $I_{0\textrm{,}g}=(c\epsilon_0/2)|B_0|^2=2P/(\pi w_g^2)$.

An axicon is a conical lens of base angle $\alpha$ made of material with a refractive index $n_{rf}$. It transforms an incident phase front $E_{in}(\rho)$ according to,
\begin{equation}
	\label{eq:Bessel12}
	E_{axicon}(\rho) = E_{in}(\rho)\exp(-ik_\rho\rho)~,
\end{equation}
where
\begin{equation}
	\label{eq:Bessel13}
	k_\rho \equiv k\sin[(n_{rf}-1)\tan\alpha]~.
\end{equation}

The product $-k_\rho\rho$ in (\ref{eq:Bessel12}) is then the phase shift suffered by a paraxial beam ray passing through the axicon at a distance $\rho$ from the optical axis.
In contrast, a thin lens with focal distance $f$ transforms a phase front according to,
\begin{equation}
	\label{eq:Bessel14}
	E_{lens}(\rho) = E_{in}(\rho)\exp(-ik\rho^2/2f)~.
\end{equation}
The transformation of a beam due to propagation in free space from position $z_0$ to position $z$ is obtained via \cite{Goodman96,Voelz11},
\begin{align}
    \label{eq:Bessel15}
	E_{prop}(x,y,z) & = H\star G\\
		& = \mathcal{F}^{-1}\{H(z)\cdot\mathcal{F}[E_{in}(x,y,z_0)]~,\nonumber
\end{align}
where $H(z)=\exp\left(ikz\sqrt{1-(\lambda k_x)^2-(\lambda k_y)^2}\right)$ and $\mathcal{F}$ denotes the two-dimensional Fourier transform in $x$ and $y$. The concatenation of the operations (\ref{eq:Bessel11}) to (\ref{eq:Bessel15}) allows to determine the radial distribution of the electric field amplitude at any location $z$ of the optical axis.

\begin{figure}[h!]
	\centerline{\includegraphics[width=8.7 truecm]{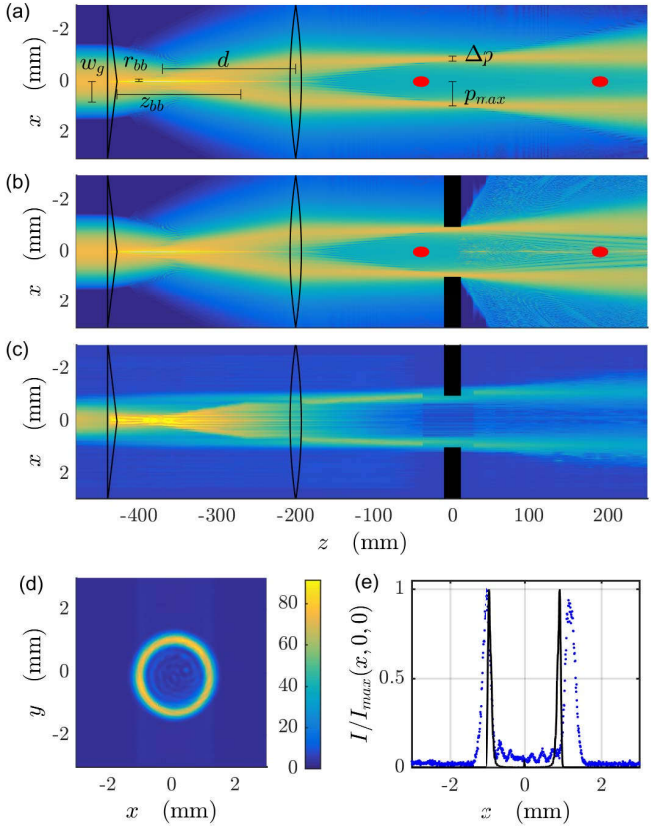}}
    \caption{(color online) Simulation of a Bessel beam produced from an incident Gaussian beam with waist radius $w_g=0.5~$mm, wavelength $\lambda=461~$nm, and power $P=10~$mW using one axicon (base angle $\alpha=0.5^\circ$, refraction index $n_{rf}=1.51$) located at $z=-440~$mm and one lens (focal length $f=200~$mm) located at $z=-200~$mm. Fig.~(a) shows the intensity distribution $I(x,y=0,z)$ in the $y=0$ plane. The colorbar codifies the intensity in a dBm scale . The positions of the 2D-MOT and the 3D-MOT are indicated by red circles. The distances indicated in Fig.(a) refer to the analytic modelling of Sec.~II.B (b) Same as (a), but in the presence of a differential pumping tube located at the position $z=0$. (c) Measurement of the evolution of a Bessel beam. Fig.~(d) shows a radial cut of the experimental intensity profile at $z=-60~$mm ; and (e) shows an axial cut of the normalized intensity profile $I(x,y=0,z=-60)$ for the experimental data (blue - dotted) and the simulation (solid line).}
    \label{fig:Propagation}
\end{figure}

We now apply this formalism to the optical system shown in Fig.~\ref{fig:Scheme}. An axicon with base angle $\alpha=0.5^\circ$ and refractive index $n_{rf}=1.51$ is placed at $z=-440~$mm on the optical axis and a converging lens with focal distance $f=200~$mm at $z=-200~$mm. The origin of the system, $z=0$, is chosen to be located in the plane, where the hollow beam intensity is maximum, that is, in the geometrical focal plane of the lens. At the same time, we place the geometric center of the funnel in this plane. A collimated Gaussian laser beam with waist radius of $w_g=0.5~$mm and wavelength $\lambda=461~$nm is injected and propagated along the optical axis using the transfer functions (\ref{eq:Bessel11}-\ref{eq:Bessel15}).
The simulation, exhibited in Fig.~\ref{fig:Propagation}(a), reveals the formation of a more than 20~cm long hollow beam near the origin which, when the light is tuned to the blue of an atomic resonance, may serve to guide cold atoms. Fig.~\ref{fig:Propagation}(b) shows the same simulation, but in the presence of a 20~cm long and 2~mm wide funnel near the origin, through which the BB is threaded. The funnel is introduced into the simulations by simply removing all radial components of the intensity distribution exceeding the funnel diameter over its whole length. A comparison between both simulations reveals the presence of diffraction caused by the funnel: even though the bulk part of the BB clearly passed through the tube, --we note a negligible loss of total light power due to the passage through the funnel of about 4\% for the chosen geometry--, peripheral partial waves blocked by the funnel lead to interference filling the interior of the tube with non-negligible light intensity.

In practice, imperfections of the axicon will spoil to some extend the quality of the intensity profile of the BB. To obtain a feeling of that, we mapped the radial intensity distribution at various positions on the optical axis using a CCD camera. The results, shown in Fig.~\ref{fig:Propagation}(c), demonstrate a good agreement.

\subsection{Higher-order Bessel-Gauss description}

The dipolar optical force exerted by a BB is obtained as the gradient of its intensity distribution. This presupposes that the intensity distribution had been numerically calculated on a sufficiently dense grid, and that the force be interpolated between grid points. This makes the Monte-Carlo simulation, to be presented in the following, heavy and slow. The calculation becomes simpler if an approximate but otherwise reliable analytical solution can be found. In this section we investigate whether the analytical description of a paraxial higher-order Bessel-Gauss (HOBG) beam can be used to model the intensity distribution generated by an axicon with a sufficient level of accuracy by comparing it with the numerical propagation of the phase front through the optical system, described in Sec.~II.A.

We again consider the optical system of Fig.~\ref{fig:Scheme}, which is composed by an axicon followed by a focusing lens. Assuming a cylindrical coordinate system centered at the geometrical focus of the lens, a paraxial HOBG of order $\nu$, transverse wavenumber $k_{\rho}$ and electric field strength $E_0$ is given by \cite{LuXH03,Gori87,Zamboni-Rached12}:
\begin{align}
	\label{eq:Bessel21}
	E_{bs}(\rho,\phi,z)  = &  ~E_0\frac{-i}{2kzq(z)}\exp\left[i\left(kz+\nu\phi+\frac{k\rho^2}{2z}\right)\right] ~\times\\
		& \times~ \exp\left[\frac{-1}{4q(z)}\left(\frac{k_\rho^2}{k^2}+\frac{\rho^2}{z^2}\right)\right]
				~J_\nu\left(\frac{ik_\rho\rho}{2kzq(z)}\right)~,\nonumber
\end{align}
where in (\ref{eq:Bessel21}) we have defined $q(z)\equiv\frac{1}{k^2w_0^2}-\frac{i}{2kz}$, in which $w_0$ is the waist radius of the HOBG. Introducing the abbreviation $u(z,\rho)=ik_\rho\rho/[2kzq(z)]$, the intensity can be written as,
\begin{equation}
	\label{eq:Bessel22}
	\begin{split}
	I_{bs}(\rho,z) = &
	 \tfrac{c\varepsilon_0}{2}|E(\rho,\phi,z)|^2\\
	= & I_0\left|\frac{1}{2kzq}\exp{\left(\frac{-k_\rho^2}{4k^2q}+\frac{k^2u^2q}{k_\rho^2}\right)}J_\nu(u)\right|^2\\
	= & \frac{I_0k^2w_0^4}{4z^2+k^2w_0^4}\exp\left(\frac{-k_\rho^2z^2-k^2\rho^2}
				{2z^2/w_0^2+k^2w_0^2/2}\right)|J_\nu(u)|^2~.
	\end{split}
\end{equation}
with $I_{0}=(c\epsilon_0/2)|E_0|^2$. The intensity is normalized as,
\begin{equation}
	\label{eq:Bessel23}
		\begin{split}
			 P_{bs} = &
			 \int_{\mathbb{R}^2}I_{bs}(\rho,0)\rho d\rho d\phi\\
			 = & I_02\pi\int_0^\infty\rho \exp{(-2\rho^2/w_0^2)}J_\nu\left(k_\rho\rho\right)^2d\rho\\
			 = & \frac{\pi w_0^2}{2}I_0\exp{(-k_\rho^2w_0^2/4)}~I_\nu(\tfrac{1}{4}k_\rho^2w_0^2)~,
		\end{split}
\end{equation}
where $I_\nu$ is the modified Bessel function of the first kind. This allows us to fix $I_0$ for a given set of parameters $P_{bs}$, $w_{0}$, $\nu$, and $k_\rho$.

By using Gaussian beam optics through a lens, the beam waist $w_0$ of the HOBG can be related with the radius of the central lobe $r_{bb}$ of the $0^{\text{th}}$ order BB formed after the axicon and with the distance $d$ between the $0^{\text{th}}$ order BB focal spot and the lens (distances indicated in Fig.~\ref{fig:Propagation}(a)).
The Bessel zone length formed after the axicon is $z_{bb}=w_g/ \tan \beta = 112~$mm \cite{Durnin87}, with $\beta=(n_{rf} -1)\tan \alpha $. Thus, the distance $d$ is calculated as $d=z_{ax\text{-}ls}-z_{bb}/2 = 184~$mm, where $z_{ax\text{-}ls}=240~$mm is the axicon-lens distance. The central lobe radius $r_{bb}$ is given by $r_{bb}=2.405/k_{\rho,bb}=39.6~\mu$m \cite{McGloin05}, with $k_{\rho,bb}=k\sin\beta$ being the transverse wavenumber of the $0^{\text{th}}$ order BB (2.405 is approximately the first non-null root of $J_0(.)$). Finally, the beam waist of the HOBG is evaluated by \cite{Haus84}:
\begin{equation}
	\label{eq:Bessel24}
	w_0 = \left[\frac{1}{r_{bb}^2}\left(1-\frac{d}{f}\right)^2+\frac{1}{f^2}\left(\frac{\pi r_{bb}}{\lambda}\right)^2\right]^{-1/2}~,
\end{equation}
which yields $w_0=0.41~$mm for our optical system.

Geometric optics predicts the radius of the ring pattern and its width at $z=0~$mm as $\rho_{max}=\beta f$ and $\Delta\rho = 3.3(\lambda f)/(\pi w_g)$, respectively \cite{Belanger78}, whose values for our optical system are $\rho_{max}=0.890~$mm and $\Delta\rho=0.194~$mm. This is in good agreement with the values found in the numerical method of Sec.~II.A.

The transverse wavenumber $k_\rho$ and the order $\nu$ of the HOBG can be determined from $\rho_{max}$ and $\Delta\rho$ values by using the following approach. The field must assume its maximum value at $\rho=\rho_{max},\phi=0,z=0$:

\begin{equation}
	\label{eq:Bessel25}
    E(\rho_{max},0,0) = E_0\exp{(-(\rho_{max}/w_0))^2)}J_\nu(k_\rho\rho_{max})~,
\end{equation}
hence, the maximum value of the function $J_\nu(s)$ must be at $s=k_\rho\rho_{max}$. Since the ring intensity decays at $\rho = \rho_{max} + \Delta\rho/2$, the first root of $J_\nu(.)$ must be equal to $k_\rho\cdot(\rho_{max} + \Delta\rho/2)$. Therefore, $\nu$ and $k_\rho$ are evaluated from the following two expressions:
\begin{align}
    \label{eq:Bessel_num_solver}
    \max[J_\nu(s)] ~~ \textrm{at} ~~ s &= k_\rho\rho_{max}~,\\
    \textrm{first root of~} J_\nu(.) & = k_\rho\cdot(\rho_{max} + \Delta\rho/2)~,\nonumber
\end{align}
which can be solved numerically. We find for our optical system $\nu=30$ and $k_\rho = 2.68~\textrm{x}~ 10^{-3} k$.

\begin{figure}[ht!]
	\centerline{\includegraphics[width=8.7cm]{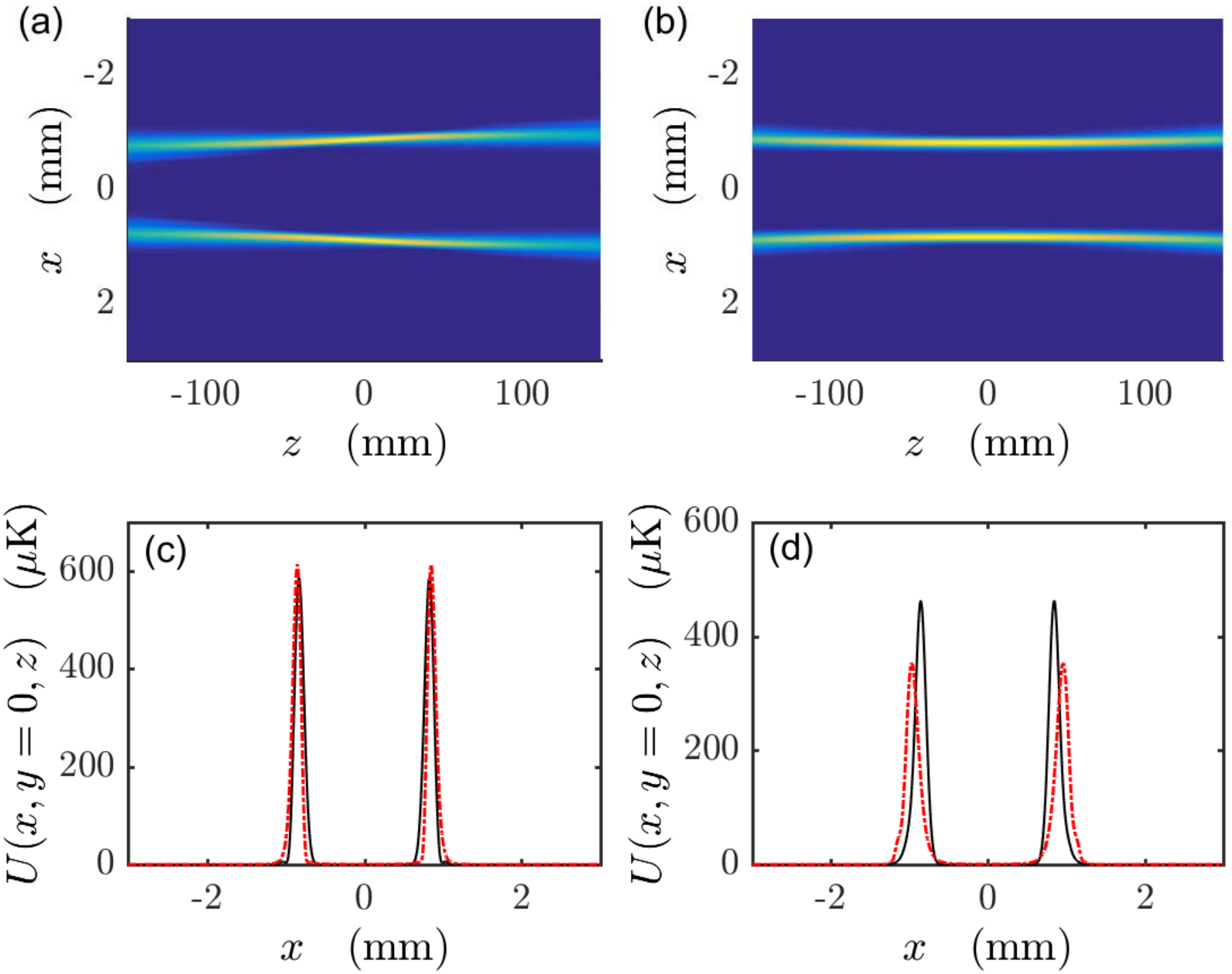}}
    \caption{(color online) (a) Intensity distribution of the hollow Bessel beam calculated by numerical propagation of the phase front through the optical system of Fig.~\ref{fig:Scheme}. (b) Same as (a), but now calculated from Eq.~(\ref{eq:Bessel21}) using the HOBG method calculated with order $\nu=30$, waist radius $w_0=0.41~$mm and transverse wavenumber $k_\rho = 2.68~\textrm{x}~ 10^{-3} k$.(c) The red dash-dotted curve shows the optical potential for a detuning of $\Delta_{bs}=10\Gamma$ calculated from a cut through the intensity profile (a) at position $z=0$. The black solid curve illustrates the expression (\ref{eq:Forces21}), see Sec.~III.C. (d) Same as (c), but at position $z=100~$mm.}
    \label{fig:CompareAnalytic}
\end{figure}

The intensity of the hollow beam calculated by the numerical propagation of the phase front is shown in Fig.~\ref{fig:CompareAnalytic}(a) whereas the intensity of the HOBG beam, calculated from Eq.~(\ref{eq:Bessel22}) with order $\nu=30$, waist radius $w_0=0.41$ mm and $k_\rho = 2.68 ~\textrm{x}~ 10^{-3} k$, is presented in Fig.~\ref{fig:CompareAnalytic}(b) in the $xz$-plane. Apparently, the absolute error is significantly larger for $z\gg 0$.

The optical dipole potential is calculated from the intensity using Eq.~(\ref{eq:Forces21}) (see Sec.~III.C). The potential produced by the numerical propagation method is represented in Figs.~\ref{fig:CompareAnalytic}(c,d) by full black lines for different values of $z$. The potential calculated via HOBG is represented in the same figures by dashed red lines. The good agreement between the potentials indicates that HOBG beams are able to describe hollow beams generated by axicons. However, the level of precision is high only for $-100~$mm $\leq$ z $\leq$ $50~$mm. The deviation between the HOBG description and the numerical propagation method tends to increase with the distance from the focal plane. For the purpose of this paper the precision of the analytical HOBG method is sufficient, so that for subsequent calculations of the forces acting on the atoms we will resort to this method.

\section{Forces acting on the atoms}

Let us now take a closer look at the layout of the proposed experiment schematized in Fig.~\ref{fig:Scheme}. The vacuum system consists of two chambers: a preparation chamber and a science chamber. In the preparation chamber strontium atoms are ejected from a heated dispenser and subject to several laser beams (i)~for trapping and cooling them in a 2D-MOT, (ii)~for pushing them via radiation pressure along the symmetry axis toward the science chamber, and (iii)~for guiding them inside a blue-detuned hollow laser beam.

To understand quantitatively how the atoms are transferred between the 2D and 3D-MOT and, in particular, the role of a hollow guiding beam, we perform simulations of atomic trajectories under the influence of the forces exerted by the laser beams and of gravity. 

As can be seen in Fig.~\ref{fig:CompareAnalytic}(c), the BB forms, near the origin and for a total power of $P_{bs}=10~$mW, a wall high enough to radially confine atoms having a kinetic energy below $k_B600~\mu$K. In a typical strontium MOT with its cloud at thermal equilibrium, however, temperatures can be as hot as 10~mK, which means that out of the Boltzmann energy distribution only a fraction of atoms will be affected by this potential barrier. Furthermore, in a MOT the atoms are also subject to a restoring force, which adds potential energy to the system. We account for this complicated situation by a simple but realistic model, assuming that the friction force exerted by the MOT is maintained at balance with a fluctuating Langevin force caused by the random photonic recoil imparted to the atoms upon light scattering. This balance is what determines the temperature of the cloud, as long as it interacts with the 2D-MOT cooling beams.

Let us now write down the forces.

\subsection{2D-MOT (i)}

The 2D-MOT generally captures atoms from a background gas or a dispenser and cools them down to the Doppler limit of the employed cooling transition. In a simple linearized approach the atomic motion [indicated by the coordinates $(\mathbf{r},\mathbf{p}$)] in a MOT can be described as a damped harmonic oscillation with a spring constant $\kappa_{mot}$ and a friction coefficient $\gamma_{mot}$,
\begin{equation}
	\label{eq:Forces01}
	\mathbf{F}_{mot}(\mathbf{r},\mathbf{p}) = -\kappa_{mot}\mathbf{r}-\gamma_{mot}\mathbf{p}~.
\end{equation}
Typical values are $\kappa_{mot}\approx 4~\textrm{x}~ 10^{-19}~$N/m and $m\gamma_{mot}\approx 2~\textrm{x}~ 10^{-22}~$Ns/m, where $m$ is the mass of the $^{88}$Sr atoms. Being subject only to these forces, the atoms would quickly cool down to zero temperature. However, the temperature is limited by spontaneous photon scattering processes leading to a random walk of the atoms in momentum space. For a cloud at thermal equilibrium at temperature $T$, using the fluctuation-dissipation theorem, we describe this diffusion by an additional stochastic Langevin force $\vec\xi(t)$, which is $\delta$-correlated \cite{Gardiner-04},
\begin{align}
	\label{eq:Forces02}
	\langle\vec\xi(t)\rangle & = 0\\
	\langle\vec\xi(t)\vec\xi(t+\tau)\rangle & = \frac{m^2}{k^2}2\gamma_{mot}^2D_T\delta(\tau)
			= 8\gamma_{mot}mk_BT\delta(\tau)~,\nonumber
\end{align}
here $D_T=\sigma^2/\gamma_{mot}$ is the diffusion coefficient and $\sigma=2k\sqrt{k_BT/m}$ the width of the Maxwell-Boltzmann distribution. $\lambda=2\pi/k=2\pi c/\omega=461~$nm is the wavelength of strong cooling transition $^1S_0$-$^1P_1$ of strontium, whose linewidth is $\Gamma=(2\pi)~30.5~\textrm{MHz}$.

In practice, we set the capture region of the 2D-MOT to $|\rho|,|z+30~\textrm{mm}|<8~$mm (left red circle in Fig.~\ref{fig:Scheme}), i.e.~the forces $\mathbf{F}_{mot}$ and $\vec\xi(t)$ are assumed to be present only within this volume.

\subsection{Push beam and gravity (ii)}

Many 2D-MOT schemes make use of an additional 'push beam', directed along the symmetry axis of the atomic cloud. Being tuned to resonance, it exerts a radiation pressure force, which can be described by \cite{metcalf2001laser},
\begin{equation}
	\label{eq:Forces11}
	\mathbf{F}_{psh}(\mathbf{p}) = \hbar k\mathbf{\hat e}_z\gamma_{psh}(\mathbf{p})~,
\end{equation}
where
\begin{equation}
	\label{eq:Forces12a}
	\gamma_{psh}(\mathbf{p}) = \frac{6\pi}{k^2}\frac{s_{psh}(\mathbf{p})}{1+s_{psh}(\mathbf{p})}
		\frac{I_{sat}}{\hbar\omega}~.
\end{equation}
is the scattering rate, which depends on the atomic velocity along the optical axis, $p_z/m$, because of the Doppler shift sensed by the accelerated atoms. The saturation parameter is given by,
\begin{equation}
	\label{eq:Forces12b}
	s_{psh}(\mathbf{p}) = \frac{I_{psh}/I_{sat}}{1+(2kp_z/m\Gamma)^2}~.
\end{equation}
Here, the push laser is assumed to be a plane wave with homogeneous intensity $I_{psh}$ limited to a radius of $\rho<1~$mm, and its goal is to accelerate the atoms toward the science chamber.

The radiation pressure force is always accompanied by heating of the atomic cloud due to the randomness of spontaneous emission. The average heating rate can be estimated by,
\begin{equation}
	\label{eq:Forces13}
	R_{psh}(\mathbf{p})  = \hbar\omega_{rec}\gamma_{psh}(\mathbf{p})~,
\end{equation}
where $\omega_{rec}=\hbar k^2/2m$ is the recoil shift.

When the transfer is too slow (for example, if the push beam intensity is too weak), the atoms deviate from the optical axis due to gravity acting perpendicularly to the optical axis:
\begin{equation}
	\label{eq:Forces14}
	\mathbf{F}_{grv} = mg\mathbf{\hat e}_x~.
\end{equation}

\subsection{Dipolar potential and radiation pressure (iii)}

If not otherwise stated, we assume a total power provided by the laser from which the BB is generated of $P_{bs}=10~\textrm{mW}$, and we tune it by $\Delta_{bs}=10\Gamma$ to the blue of the cooling transition. The dipolar potential sensed by the atoms is then,

\begin{equation}
	\label{eq:Forces21}
	U_{bs}(\mathbf{r}) =\frac{\hbar\Delta_{bs}}{2}\ln\left[1+s_{bs}(\mathbf{r})\right]
\end{equation}
where we defined the local velocity-dependent saturation parameter
\begin{equation}
	s_{bs}(\mathbf{r}) = \frac{I_{bs}(\mathbf{r})/I_{sat}}{1+(2\Delta_{bs}/\Gamma)^2}
\end{equation}
with $I_{sat}=2\pi^2c\hbar\Gamma/(3\lambda^3)$ being the saturation intensity and $\int_{R^3}I_{bs}(\mathbf{r})dxdy=P_{bs}$. The inhomogeneity of this potential gives rise to a conservative force,

\begin{equation}
    \label{eq:Forces22}
   \mathbf{F}_{bs}(\mathbf{r}) = -\nabla U_{bs}(\mathbf{r})~.
\end{equation}
On the other hand, when the laser is not tuned very far from atomic resonance, the atoms will also sense a radiation pressure force oriented in the propagation direction of the laser beam and being proportional to the local scattering rate,
\begin{equation}
    \gamma_{bs}(\mathbf{r}) = \frac{6\pi}{k^2}\frac{s_{bs}(\mathbf{r})}{1+s_{bs}(\mathbf{r})}\frac{I_{sat}}{\hbar\omega}~.
\end{equation}
The radiation pressure force can be expressed by
\begin{equation}
    \label{eq:Forces25}
   \mathbf{F}_{rp}(\mathbf{r}) = \hbar k\mathbf{\hat e}_z\gamma_{bs}(\mathbf{r})~.
\end{equation}

Again, spontaneous emission from the Bessel beams lead to heating estimated by the rate,
\begin{equation}
	\label{eq:Forces26}
	R_{bs}(\mathbf{r}) = \omega_{rec}\gamma_{bs}(\mathbf{r})~.
\end{equation}

The intensity distribution of the BB, which is required to determine the forces from Eq.~(\ref{eq:Forces22}) and Eq.~(\ref{eq:Forces25}) is calculated numerically, as explained in Sec.~II.A, or approximated analytically from the expression (\ref{eq:Bessel22}). In the appendix we also derive an analytical expression for the dipolar force.

\subsection{Implementation of the Monte-Carlo simulation}

The Langevin equations for the motion of $N$ (non-interacting) atoms (labeled by an index $n$) subject to the light forces exerted by the BB, the resonant push beam, and the MOT beams are now,
\begin{align}
    \label{Forces31}
   \mathbf{\dot r}_n(t)&  = \tfrac{1}{m}\mathbf{p}_n(t)\\
		\mathbf{\dot p}_n(t) & = \mathbf{F}(\mathbf{r}_n,\mathbf{v}_n) 
			= \mathbf{F}_{mot}(\mathbf{r}_n,\mathbf{p}_n)+\mathbf{F}_{psh}(\mathbf{p}_n)\nonumber\\
			& \qquad +\mathbf{F}_{grv}+\mathbf{F}_{bs}(\mathbf{r}_n)+\mathbf{F}_{rp}(\mathbf{r}_n)+\vec\xi_n(t)~.\nonumber
\end{align}
We implement the simulation of the atomic motion by a stochastic Runge-Kutta algorithm \cite{Honeycutt92}. The simulations have been carried out in two dimensions within the $y=0$ plane,
\begin{equation}
    \label{eq:Forces32}
   \begin{array}{rcrcl}
		\mathbf{r}_n(dt) & = & \mathbf{r}_n^{(0)} & + & \tfrac{dt}{2m}[\mathbf{p}_n^{(0)}+\mathbf{p}_n^{(1)}] \vspace{.2cm}\\
		\mathbf{p}_n(dt) & = & \mathbf{p}_n^{(0)} & + & \tfrac{dt}{2}[\mathbf{F}(\mathbf{r}_n^{(0)},\mathbf{p}_n^{(0)})
				+\mathbf{F}(\mathbf{r}_n^{(1)},\mathbf{v}_n^{(1)})] \vspace{.2cm}\\
			& & & + & \vec\zeta_n\sqrt{8\gamma_{mot}mk_BTdt}
	\end{array}
\end{equation}
with
\begin{equation}
	\label{Forces33}
    \begin{array}{r c l}
       \mathbf{r}_n^{(1)} & = &\mathbf{r}_n^{(0)}+\frac{1}{m}\mathbf{p}_n^{(0)}dt \vspace{.2cm}\\
		\mathbf{p}_n^{(1)} & = & \mathbf{p}_n^{(0)}+\mathbf{F}(\mathbf{r}_n^{(0)},\mathbf{p}_n^{(0)})dt
		+\vec\zeta_n\sqrt{8\gamma_{mot}mk_BTdt}~.\nonumber
	\end{array}
\end{equation}

The Cartesian components of $\zeta$ are $\delta$-correlated random numbers satisfying $\langle\zeta_n\rangle=0$ and $\langle\zeta_n(t)\zeta_m(t+\tau\rangle=\delta_{mn}\delta(\tau)$.

Heating of the atomic cloud by spontaneous emission from the push and the Bessel beam is incorporated by the following procedure. After each time step of the simulation ($dt=10$ $\mu$s) a direction $\mathbf{k}_{psh,n}$ is randomly generated for every atom, accounting for the randomness of the direction into which the light from the push beam is scattered. The same procedure is repeated for the Bessel beam, $\mathbf{k}_{bs,n}$. The momentum change due to recoil is calculated for every atom by,
\begin{equation}
    \begin{array}{r c l}
	\label{Forces35}
	\mathbf{p}_{n}(t+dt) = \mathbf{p}_{n}(t) & + & \mathbf{k}_{psh,n}\sqrt{2mR_{psh}(\mathbf{p}_{n})dt}\\
		& + & \mathbf{k}_{bs,n}\sqrt{2mR_{bs}(\mathbf{r}_{n})dt}~.\nonumber
	\end{array}
\end{equation}
We verified that this procedure generates the expected temperature increase due to heating.

Atoms hitting the walls of the funnel located at $|\rho|>1~\textrm{mm}$ and $|z|<10~\textrm{mm}$ are removed from the simulation. Only atoms passing through the capture region of the 3D-MOT, which is set to $|\rho|,|z-190~\textrm{mm}|<3~$mm, contribute to the transfer efficiency $\eta\equiv N_{trp}/N$, where $N_{trp}$ is the number of recaptured atoms. Finally, the transfer process is aborted after 20~ms, time after which atoms still flying around in the vacuum chamber are assumed not to make it to the 3D-MOT region.

\subsection{Results of the simulation}

In order to evaluate the impact of the BB on the transfer efficiency from the 2D to the 3D-MOT, we run simulations of the Eqs.~(\ref{eq:Forces32}) with $N=1,000$ atoms assumed to be at equilibrium in the 2D-MOT at temperatures between $T=1~$mK and $100~$mK, and we vary the intensity of the BB and the push beam. We verified that the number of simulated atoms is sufficient to generate reproducible dependencies. In the simulations, we let the 2D-MOT equilibrate at this temperature before switching on the push and BBs. We noticed that, the simultaneous presence of a strong BB can shield the central 2D-MOT region, reduce its loading efficiency, and thus spoil the transfer efficiency of the 3D-MOT.

\begin{figure}[ht]
	\centerline{\includegraphics[width=8.7cm]{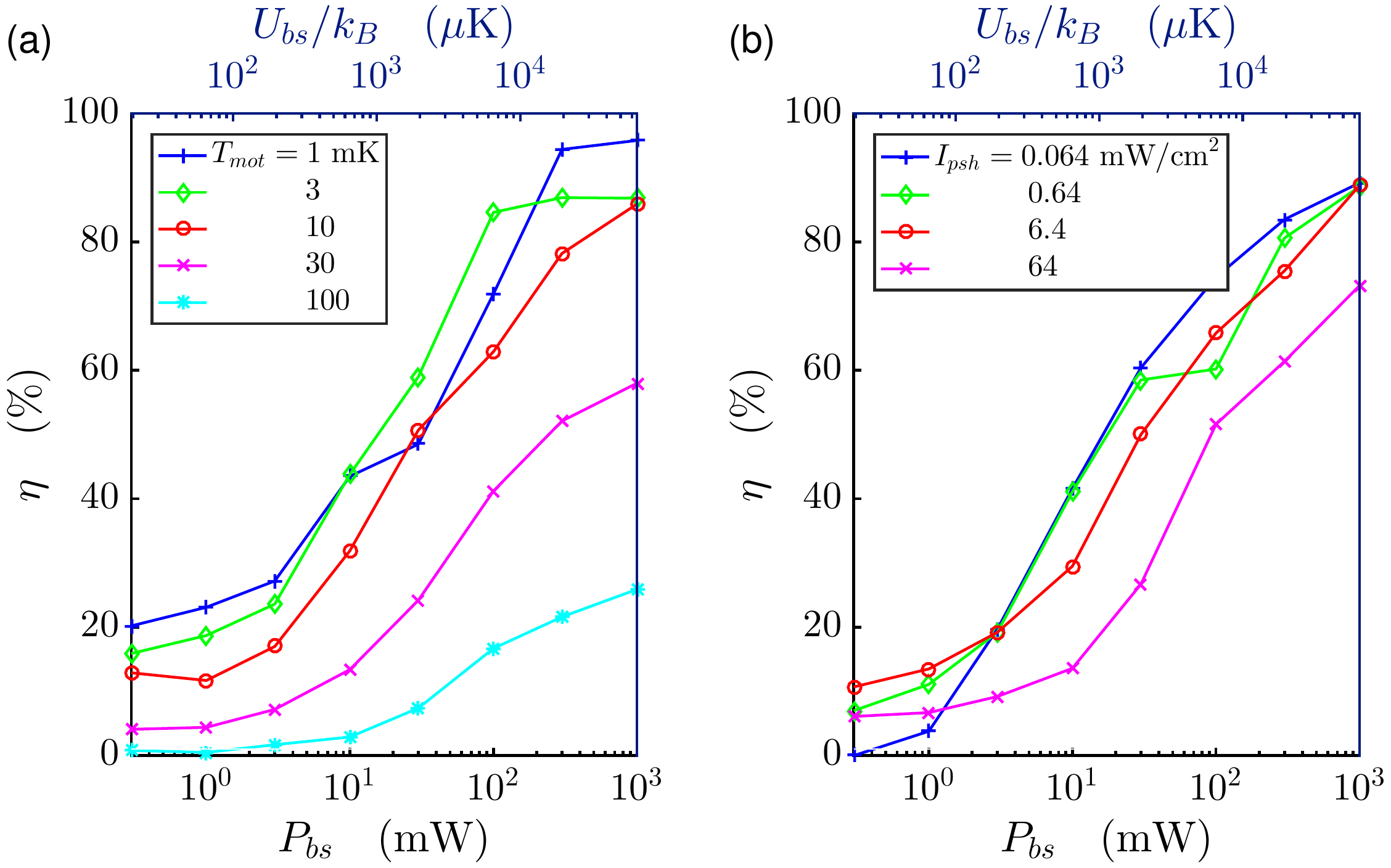}}
    \caption{(color online) (a) Simulation of the loading efficiency as a function of the potential depth $U_{0,bs}$ 
		of the Bessel beam for various temperatures between $T=1~$mK and $100~$mK and a fixed push beam intensity set to 
		$I_{psh}=6.4~\textrm{mW/cm}^2$.
		(b) Same as (a), but now for various push beam intensities between $I_{psh}=0.064~\textrm{mW/cm}^2$ and 
		$64~\textrm{mW/cm}^2$, while the temperature is set to $T=10~$mK.}
    \label{fig:HeatBess}
\end{figure}

The results are shown in Figs.~\ref{fig:HeatBess}, which reproduce the number of recaptured atoms as a function of various parameters. The curves of Fig.~\ref{fig:HeatBess}(a) demonstrate clearly that higher BB intensities help to guide the atoms and thus to enhance the transfer efficiency. This is particularly important, when the temperature of the atomic cloud is high. With the chosen beam geometry and for a laser detuning of $\Delta_{bs}=10\Gamma$ total beam powers higher than $P_{bes}=10~$mW seem to be necessary. Above 100~mW beam power an increase of the transfer efficiency from below 20\% to more than 80\% can be expected for an atomic cloud as hot as $T=10~$mK.

At $T=10~$mK the mean radial velocity of the atoms in the 2D-MOT is $\bar v=\sqrt{k_BT/m}=1~$m/s. This is by far too slow for the atoms to reach the 3D-MOT, which is distant by 23~cm. By the time atoms with this velocity arrive at the recapture region via ballistic flight, they will have suffered a gravitational sag of more than 20~cm! The benefit of pushed atomic transfer has been confirmed by experimental observations \cite{Nosske17}. A resonant push beam with an intensity of $I_{psh}=6.4~\textrm{mW/cm}^2$ is able to accelerate the atoms to 30~m/s by the time they arrive at the 3D-MOT, despite the fact that the acceleration decreases as soon as the Doppler shift exceeds the natural linewidth. This reduces the gravitational sag to only a few $10~\mu$m. The BB now helps to defeat gravity, i.e.~even at very weak push beam intensities atoms may slowly drifts across the hollow BB toward the 3D-MOT. This explains the increase of the loading efficiency observed in Fig.~\ref{fig:HeatBess}(b) even for very low push beam intensities of below $0.6~\textrm{mW/cm}^2$. Higher push beam intensities decrease the transfer time. However, at intensities exceeding the saturation intensity, $I_{sat}=43~\textrm{mW/cm}^2$, heating starts to spoil the transfer efficiency, as verified in Fig.~\ref{fig:HeatBess}(b). We point out that the curve of Fig. Fig.~\ref{fig:HeatBess}(b) corresponding to $64~\textrm{mW/cm}^2$ must be taken as an upper bound, because other effects, such as the perturbation of the 3D-MOT by resonant push light is not captured by our model.
The radiation pressures exerted by the BB and heating are included in the simulation via the expressions~(\ref{eq:Forces25}) and (\ref{eq:Forces26}). At small detunings, radiation pressure assists the transfer by accelerating the atoms toward the science chamber. On the other hand, heating tends to increase the transverse velocity components and cause atomic losses. However, being guided by the dark hollow channel of the BB, the atoms spend only little time exposed to the BB light, so that heating is limited.

\section{Conclusion and perspectives}

We presented and evaluated numerically, as well as analytically, a very simple scheme for enhancing the loading efficiency of a 3D magneto-optical trap by guiding a precooled atomic cloud from a preparation chamber through a differential pumping tube to a science chamber via a hollow BB. We quantitatively verified the intuition that higher temperatures of the precooled cloud require a higher potential depth of the BB. At the example of $^{88}$Sr atoms precooled to 10~mK, we found that it should be feasible to increase the loading efficiency from 20\% to 80\% using, e.g.,~a BB of total power $P_{bs}=500~$mW tuned $50\Gamma$ above the atomic transition at 461~nm. While the obtained results are specific for the considered experimental scheme, the general procedure of the simulations can easily be adapted to other schemes and geometries.

A problem may occur when trying to thread the BB through a narrow differential pumping tube: Even when the beam clearly passes through the tube, the external partial waves blocked by the tube lead to interference filling the interior of the tube with non-negligible light intensity. In our case, this does not represent a major problem, as it merely leads to a slight increase in the radiation pressure accelerating the atoms towards the 3D-MOT. Nevertheless, it should be noted that this interference may be reduced by reducing the BB diameter, which, however, requires axicons with smaller base angle, $\alpha<0.5^\circ$, or inverted axicons, i.e.~conical lenses with radially diminishing thickness. Both are not readily available on the market. However, such configurations can be realized with spatial light modulators (SLM). An interesting extension of our waveguiding scheme could be the use of recently discovered non-diffracting superpositions of BBs called "Frozen Waves" \cite{Zamboni-Rached04,Ambrosio15,Ambrosio15b,Ambrosio15c}. These are stationary localized wave fields with high transverse localization and a longitudinal intensity pattern which can assume any desired shape within long distances

\section*{Acknowledgments}

The authors acknowledge the Brazilian agencies for financial support. Ph.W.C. hold grants from The S\~ao Paulo Research Foundation FAPESP (2013/04162-5) and CAPES (88881.1439362017-01 and N88887.1301972017-01). L.A.A. thanks FAPESP (2017/10445-0) and The Brazilian National Council for Scientific and Technological Development (CNPq) (426990/2018-8 and 307898/2018-0) for supporting this work.  D.R., V.S.A., C.B. and M. M. Thank the Coordenação de Aperfeiçoamento de Pessoal de Nível Superior – Brasil (CAPES) – Finance Code 001 for the scholarships on which this study was partly financed.

\section{Disclosures}
\medskip
\noindent\textbf{Disclosures.} The authors declare no conflicts of interest.

\bibliographystyle{plain}

\begin{thebibliography}{10}

\bibitem{Ambrosio15c}
Leonardo~Andr{\'{e}} Ambrosio and Mariana {de Matos Ferreira}.
\newblock {Time-average forces over Rayleigh particles by superposition of
  equal-frequency arbitrary-order Bessel beams}.
\newblock {\em Journal of the Optical Society of America B}, 32(5):B67, may
  2015.

\bibitem{Ambrosio15b}
Leonardo~Andr{\'{e}} Ambrosio and Michel Zamboni-Rached.
\newblock {Analytical approach of ordinary frozen waves for optical trapping
  and micromanipulation}.
\newblock {\em Applied Optics}, 54(10):2584, apr 2015.

\bibitem{Ambrosio15}
Leonardo~Andr{\'{e}} Ambrosio and Michel Zamboni-Rached.
\newblock {Optical forces experienced by arbitrary-sized spherical scatterers
  from superpositions of equal-frequency Bessel beams}.
\newblock {\em Journal of the Optical Society of America B}, 32(5):B37, may
  2015.

\bibitem{Arlt02}
Jochen Arlt, Kishan Dholakia, Josh Soneson, and Ewan~M. Wright.
\newblock {Optical dipole traps and atomic waveguides based on Bessel light
  beams}.
\newblock {\em Physical Review A}, 63(6):063602, may 2001.

\bibitem{Bagnato87}
V.~S. Bagnato, G.~P. Lafyatis, A.~G. Martin, E.~L. Raab, R.~N. Ahmad-Bitar, and
  D.~E. Pritchard.
\newblock {Continuous Stopping and Trapping of Neutral Atoms}.
\newblock {\em Physical Review Letters}, 58(21):2194--2197, may 1987.

\bibitem{Bajcsy11}
M.~Bajcsy, S.~Hofferberth, T.~Peyronel, V.~Balic, Q.~Liang, A.~S. Zibrov,
  V.~Vuletic, and M.~D. Lukin.
\newblock {Laser-cooled atoms inside a hollow-core photonic-crystal fiber}.
\newblock {\em Physical Review A}, 83(6):063830, jun 2011.

\bibitem{Barrett01}
M.~D. Barrett, J.~A. Sauer, and M.~S. Chapman.
\newblock {All-Optical Formation of an Atomic Bose-Einstein Condensate}.
\newblock {\em Physical Review Letters}, 87(1):010404, jun 2001.

\bibitem{Belanger78}
Pierre-Andr{\'{e}} B{\'{e}}langer and Marc Rioux.
\newblock {Ring pattern of a lens–axicon doublet illuminated by a Gaussian
  beam}.
\newblock {\em Applied Optics}, 17(7):1080, apr 1978.

\bibitem{Voelz11}
J~B Breckinridge and D~G Voelz.
\newblock {\em {Computational Fourier Optics: A MATLAB Tutorial}}.
\newblock SPIE Press monograph. SPIE Press, 2011.

\bibitem{Brzobohaty08}
Oto Brzobohat{\'{y}}, Tom{\'{a}}{\v{s}} Ci{\v{z}}m{\'{a}}r, and Pavel
  Zem{\'{a}}nek.
\newblock {High quality quasi-Bessel beam generated by round-tip axicon}.
\newblock {\em Optics Express}, 16(17):12688, aug 2008.

\bibitem{Carrat14}
Vincent Carrat, Citlali Cabrera-Guti\'{e}rrez, Marion Jacquey, Jos\'{e}~W.
  Tabosa, Bruno~Viaris de~Lesegno, and Laurence Pruvost.
\newblock Long-distance channeling of cold atoms exiting a 2d magneto-optical
  trap by a laguerre–gaussian laser beam.
\newblock {\em Opt. Lett.}, 39(3):719--722, Feb 2014.

\bibitem{Depret02}
Benoît D{\'{e}}pret, Philippe Verkerk, and Daniel Hennequin.
\newblock {Characterization and modelling of the hollow beam produced by a real
  conical lens}.
\newblock {\em Optics Communications}, 211(1-6):31--38, oct 2002.

\bibitem{Dieckmann98}
K.~Dieckmann, R.~J.~C. Spreeuw, M.~Weidem{\"{u}}ller, and J.~T.~M. Walraven.
\newblock {Two-dimensional magneto-optical trap as a source of slow atoms}.
\newblock {\em Physical Review A}, 58(5):3891--3895, nov 1998.

\bibitem{Durnin87}
J.~Durnin, J.~J. Miceli, and J.~H. Eberly.
\newblock {Diffraction-free beams}.
\newblock {\em Physical Review Letters}, 58(15):1499--1501, apr 1987.

\bibitem{Fatemi06}
F.~K. Fatemi and M.~Bashkansky.
\newblock {Cold atom guidance using a binary spatial light modulator}.
\newblock {\em Optics Express}, 14(4):1368, feb 2006.

\bibitem{Fortagh98}
J.~Fortagh, A.~Grossmann, T.~W. H{\"{a}}nsch, and C.~Zimmermann.
\newblock {Fast loading of a magneto-optical trap from a pulsed thermal
  source}.
\newblock {\em Journal of Applied Physics}, 84(12):6499--6501, dec 1998.

\bibitem{Friedman02}
Nir Friedman, Ariel Kaplan, and Nir Davidson.
\newblock Dark optical traps for cold atoms.
\newblock volume~48 of {\em Advances In Atomic, Molecular, and Optical
  Physics}, pages 99 -- 151. Academic Press, 2002.

\bibitem{Gardiner-04}
C.~Gardiner, P.~Zoller, and P.~Zoller.
\newblock {\em Quantum Noise: A Handbook of Markovian and Non-Markovian Quantum
  Stochastic Methods with Applications to Quantum Optics}.
\newblock Springer Series in Synergetics. Springer, 2004.

\bibitem{Goodman96}
J~W Goodman.
\newblock {\em {Introduction to Fourier Optics}}.
\newblock McGraw-Hill physical and quantum electronics series. W. H. Freeman,
  2005.

\bibitem{Gori87}
F.~Gori, G.~Guattari, and C.~Padovani.
\newblock {Bessel-Gauss beams}.
\newblock {\em Optics Communications}, 64(6):491--495, dec 1987.

\bibitem{Haensel01}
W.~H{\"{a}}nsel, P~Hommelhoff, T.~W. H{\"{a}}nsch, and J~Reichel.
\newblock {Bose–Einstein condensation on a microelectronic chip}.
\newblock {\em Nature}, 413(6855):498--501, oct 2001.

\bibitem{Haus84}
H~A Haus.
\newblock {\em {Waves and fields in optoelectronics}}, volume 897 of {\em
  Prentice-Hall Series in Solid State Physical Electronics}.
\newblock Prentice Hall, Incorporated, 1984.

\bibitem{Honeycutt92}
Rebecca~L. Honeycutt.
\newblock {Stochastic Runge-Kutta algorithms. I. White noise}.
\newblock {\em Physical Review A}, 45(2):600--603, jan 1992.

\bibitem{Kulin01}
S.~Kulin, S.~Aubin, S.~Christe, B.~Peker, S.~L. Rolston, and L.~A. Orozco.
\newblock {A single hollow-beam optical trap for cold atoms}.
\newblock {\em Journal of Optics B: Quantum and Semiclassical Optics},
  3(6):353--357, dec 2001.

\bibitem{LuXH03}
Xuan~Hui Lu, Xu~Min Chen, Lei Zhang, and Da~Jian Xue.
\newblock {High-Order Bessel-Gaussian Beam and its Propagation Properties}.
\newblock {\em Chinese Physics Letters}, 20(12):2155--2157, 2003.

\bibitem{Manek98}
I.~Manek, Yu.B Ovchinnikov, and R.~Grimm.
\newblock {Generation of a hollow laser beam for atom trapping using an
  axicon}.
\newblock {\em Optics Communications}, 147(1-3):67--70, feb 1998.

\bibitem{McGloin05}
D~McGloin and Kishan Dholakia.
\newblock {Bessel beams: Diffraction in a new light}.
\newblock {\em Contemporary Physics}, 46(1):15--28, jan 2005.

\bibitem{McLeod54}
John~H. McLeod.
\newblock {The Axicon: A New Type of Optical Element}.
\newblock {\em Journal of the Optical Society of America}, 44(8):592, aug 1954.

\bibitem{Mestre10}
M.~Mestre, F.~Diry, B.~{Viaris de Lesegno}, and L.~Pruvost.
\newblock {Cold atom guidance by a holographically-generated Laguerre-Gaussian
  laser mode}.
\newblock {\em The European Physical Journal D}, 57(1):87--94, mar 2010.

\bibitem{metcalf2001laser}
H.J. Metcalf and P.~van~der Straten.
\newblock {\em Laser Cooling and Trapping}.
\newblock Graduate Texts in Contemporary Physics. Springer New York, 2001.

\bibitem{Myatt96}
C.~J. Myatt, N.~R. Newbury, R.~W. Ghrist, S.~Loutzenhiser, and C.~E. Wieman.
\newblock {Multiply loaded magneto-optical trap}.
\newblock {\em Optics Letters}, 21(4):290, feb 1996.

\bibitem{Nosske17}
Ingo Nosske, Luc Couturier, Fachao Hu, Canzhu Tan, Chang Qiao, Jan Blume, Y.~H.
  Jiang, Peng Chen, and Matthias Weidem\"uller.
\newblock Two-dimensional magneto-optical trap as a source for cold strontium
  atoms.
\newblock {\em Phys. Rev. A}, 96:053415, Nov 2017.

\bibitem{Pechkis12}
Joseph~A. Pechkis and Fredrik~K. Fatemi.
\newblock {Cold atom guidance in a capillary using blue-detuned, hollow optical
  modes}.
\newblock {\em Optics Express}, 20(12):13409, jun 2012.

\bibitem{Phillips82}
William~D. Phillips and Harold Metcalf.
\newblock {Laser Deceleration of an Atomic Beam}.
\newblock {\em Physical Review Letters}, 48(9):596--599, mar 1982.

\bibitem{Poulin11}
Jerome Poulin, Philip~S. Light, Raman Kashyap, and Andre~N. Luiten.
\newblock {Optimized coupling of cold atoms into a fiber using a blue-detuned
  hollow-beam funnel}.
\newblock {\em Physical Review A}, 84(5):053812, nov 2011.

\bibitem{Prevedelli99}
M.~Prevedelli, F.~Cataliotti, E.~Cornell, J.~Ensher, C.~Fort, L.~Ricci,
  G.~Tino, and M.~Inguscio.
\newblock {Trapping and cooling of potassium isotopes in a
  double-magneto-optical-trap apparatus}.
\newblock {\em Physical Review A}, 59(1):886--888, jan 1999.

\bibitem{Prodan85}
John Prodan, Alan Migdall, William~D. Phillips, Ivan So, Harold Metcalf, and
  Jean Dalibard.
\newblock {Stopping Atoms with Laser Light}.
\newblock {\em Physical Review Letters}, 54(10):992--995, mar 1985.

\bibitem{Rhodes06}
D.~P. Rhodes, D.~M. Gherardi, J.~Livesey, D.~McGloin, H.~Melville,
  T.~Freegarde, and K.~Dholakia.
\newblock {Atom guiding along high order Laguerre–Gaussian light beams formed
  by spatial light modulation}.
\newblock {\em Journal of Modern Optics}, 53(4):547--556, mar 2006.

\bibitem{Rhodes02}
D.P. Rhodes, G.P.T. Lancaster, J.~Livesey, D.~McGloin, J.~Arlt, and
  K.~Dholakia.
\newblock {Guiding a cold atomic beam along a co-propagating and oblique hollow
  light guide}.
\newblock {\em Optics Communications}, 214(1-6):247--254, dec 2002.

\bibitem{SongY99}
Y~Song, D~Milam, and W.~T. Hill.
\newblock {Long, narrow all-light atom guide}.
\newblock {\em Optics Letters}, 24(24):1805, dec 1999.

\bibitem{Tsai06}
Tracy Tsai, Euan McLeod, and Craig~B. Arnold.
\newblock {Generating Bessel beams with a tunable acoustic gradient index of
  refraction lens}.
\newblock In Kishan Dholakia and Gabriel~C. Spalding, editors, {\em Optics
  Letters}, volume~31, page 63261F, aug 2006.

\bibitem{ZhaoyingWang05}
Zhaoying Wang, Yiming Dong, and Qiang Lin.
\newblock {Atomic trapping and guiding by quasi-dark hollow beams}.
\newblock {\em Journal of Optics A: Pure and Applied Optics}, 7(3):147--153,
  mar 2005.

\bibitem{XinyeXu01}
Xinye Xu, Kihwan Kim, Wonho Jhe, and Namic Kwon.
\newblock {Efficient optical guiding of trapped cold atoms by a hollow laser
  beam}.
\newblock {\em Physical Review A}, 63(6):063401, may 2001.

\bibitem{Zamboni-Rached04}
Michel Zamboni-Rached.
\newblock {Stationary optical wave fields with arbitrary longitudinal shape by
  superposing equal frequency Bessel beams: Frozen Waves}.
\newblock {\em Optics Express}, 12(17):4001, 2004.

\bibitem{Zamboni-Rached12}
Michel Zamboni-Rached, Erasmo Recami, and Massimo Balma.
\newblock {Simple and effective method for the analytic description of
  important optical beams when truncated by finite apertures}.
\newblock {\em Applied Optics}, 51(16):3370, jun 2012.

\end{thebibliography}

\section{Appendix}

\subsection{Calculation of the dipolar force exerted by a Bessel-Gauss beam}

In order to obtain the dipolar force exerted by the Bessel-Gauss beam, we need to take the gradient of the intensity distribution in Eq. (\ref{eq:Bessel22}),
\begin{equation}
    \label{eq:App01}
    \nabla I_{bs} = \mathbf{\hat e}_\rho\frac{\partial I_{bs}}{\partial\rho}+\mathbf{\hat e}_\phi\frac{1}{\rho}\frac{\partial I_{bs}}{\partial\phi}+\mathbf{\hat e}_z\frac{\partial I_{bs}}{\partial z} \simeq \mathbf{\hat e}_\rho\frac{\partial I_{bs}}{\partial\rho}~,
\end{equation}
neglecting $\partial_zI$. The radial gradient is hence,

\begin{equation}
 \label{eq:App02}
\begin{split}
    \frac{\partial I_{bs}}{\partial\rho}=& \frac{c\varepsilon_0E_0^2}{2}\left(\frac{1}{2kz|q|}\right)^2\exp\left(2\textrm{Re}\left(\frac{-k_\rho^2}{4k^2q}+\frac{k^2u^2q}{k_\rho^2}\right)\right)\\
    &\times\Bigg\{2J_\nu(u)J_\nu(u^*)\frac{\partial}{\partial\rho}\textrm{Re}\frac{k^2u^2q}{k_\rho^2} +J_\nu(u)\frac{\partial u^*}{\partial\rho}\frac{\partial J_\nu(u^*)}{\partial u^*}\\
    &+J_\nu^*(u)\frac{\partial u}{\partial\rho}\frac{\partial J_\nu(u)}{\partial u}\Bigg\}~,
\end{split}
\end{equation}
where we used $J_\nu^*(u)=J_\nu(u^*)$ for holomorphic functions. Using the recursion relation $\frac{\partial J_\nu(u)}{\partial u} = \frac{1}{2}\left[J_{\nu-1}(u)-J_{\nu+1}(u)\right]$ we find,
\begin{equation}
 \label{eq:App06}
\begin{split}
	\frac{\partial I_{bs}}{\partial\rho}=& \frac{c\varepsilon_0E_0^2}{2}\left(\frac{1}{2kz|q|}\right)^2\exp{\left[2\textrm{Re}\left(\frac{-k_\rho^2}{4k^2q}
	 +\frac{k^2u^2q}{k_\rho^2}\right)\right]}\\
	 &\times\Bigg\{2J_\nu(u)J_\nu(u^*)\frac{\partial}{\partial\rho}\textrm{Re}\frac{-\rho^2}{4z^2q}-\frac{k_\rho}{2kz}\textrm{Im}\Bigg[\frac{1}{q}J_\nu^*(u)\\
	 &\times\left[J_{v-1}(u)-J_{v+1}(u)\right]\Bigg]\Bigg\}\\
	 =& -\frac{c\varepsilon_0E_0^2}{2}\frac{k^4w_0^2}{(2z^2/w_0^2+k^2w_0^2/2)^2}\exp{\left(\frac{-k_\rho^2z^2-k^2\rho^2}{2z^2/w_0^2+k^2w_0^2/2}\right)}\\
	 &\times\Bigg\{\rho|J_\nu(u)|^2+\frac{1}{2}kw_0^2k_\rho z\textrm{Im~}q^*J_\nu^*(u)\\
	 &\times\left[J_{v-1}(u)-J_{v+1}(u)\right]\Bigg\}.
\end{split}
\end{equation}

\begin{figure}[h]
	\centerline{\includegraphics[width=8.7cm]{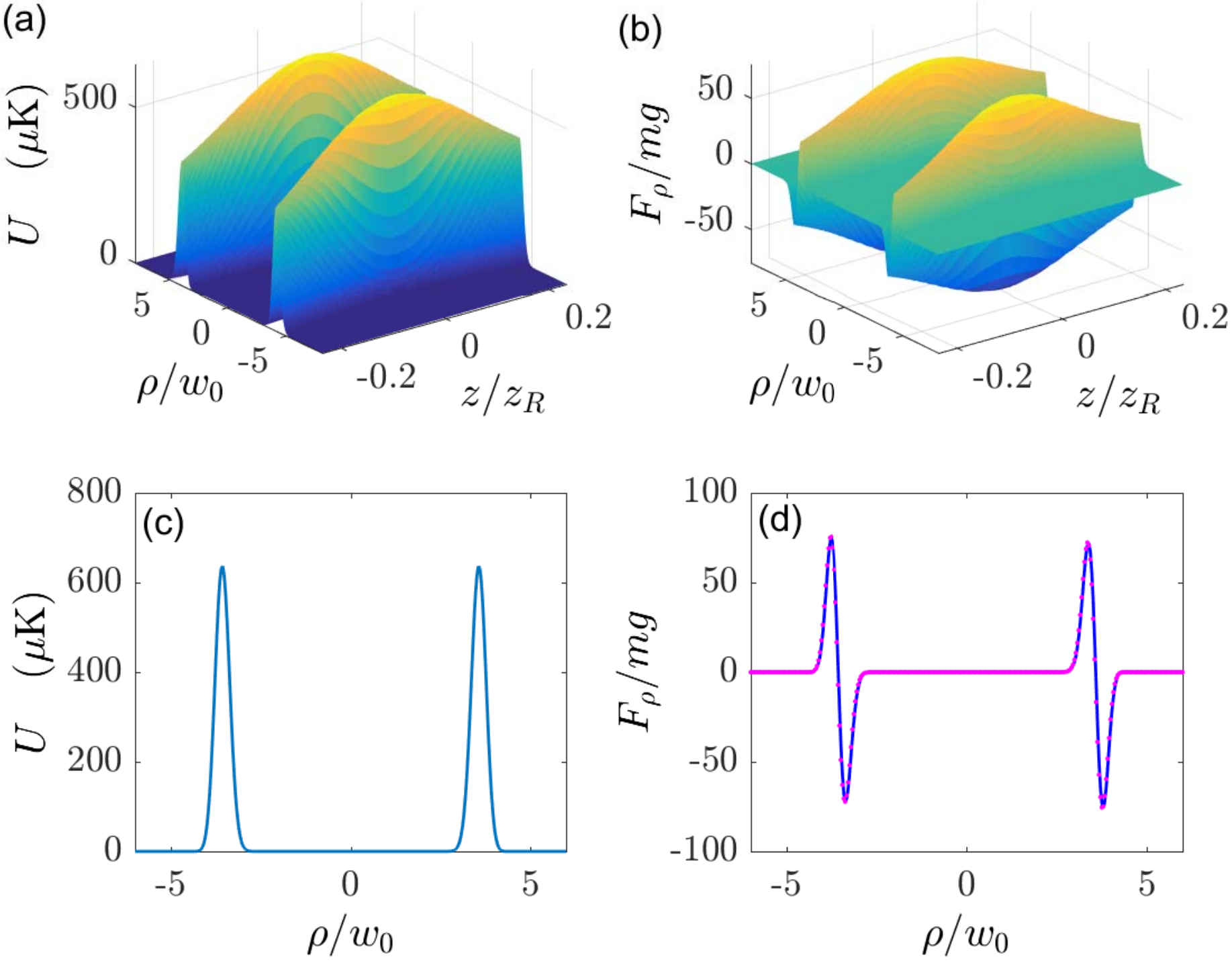}}
    \caption{(color online) Calculation of the dipolar potential (a,c) and force (b,d) exerted by a Bessel-Gauss beam 
		of order $\nu=30$ at the Fourier plane $z=0$ for a beam power of $P_{bs}=10~$mW, a waist radius of $w_0=0.41~$mm and transverse wavenumber $k_\rho = 2.68~\textrm{x}~ 10^{-3} k$. 
		The blue solid line in (d) represents the numerical gradient of the potential in (c).
		The pink dotted line is a calculation of the force according to Eq. (\ref{eq:App03}).}
    \label{fig:ForceAnalytic}
\end{figure}

Finally, the radial force exerted by the Bessel-Gauss beam is according to Eq.~(\ref{eq:Forces22}),
\begin{equation}
    \label{eq:App03}
\begin{split}
    F_\rho =& -\frac{\partial U_{bs}}{\partial\rho} = -\frac{\hbar \Delta_{bs}}{2}\frac{1}{1+s(\mathbf{r})}\frac{1}{1+(2\Delta_{bs}/\Gamma)^2}\frac{1}{I_{sat}}\frac{\partial I_{bs}(\mathbf{r})}{\partial\rho}    \\
		&\underset{|\Delta|\gg\Gamma}{\simeq}-\frac{3\pi c^2}{2\omega^3}\frac{\Gamma}{\Delta_{bs}}\frac{1}{1+s(\mathbf{r})}\frac{\partial I_{bs}(\mathbf{r})}{\partial\rho}~.
\end{split}
\end{equation}
Fig.~\ref{fig:ForceAnalytic} shows the forces calculated from Eq.~(\ref{eq:App06}).
Obviously, the forces calculated in Figs.~\ref{fig:ForceAnalytic}(b,d) are large enough to compensate for gravity.

\end{document}